\documentclass[prl,showpacs,superscriptaddress,floatfix,twocolumn]{revtex4-1}
\usepackage{epsfig}
\usepackage{amsmath}
\usepackage{amssymb}
\usepackage{amsfonts}
\usepackage{dcolumn}
\usepackage{eucal}
\usepackage{bm}
\usepackage{color}
\usepackage[colorlinks,linkcolor=blue,citecolor=blue]{hyperref}

\usepackage{epstopdf}

\usepackage{graphicx}

\def\Bcalv{\bm{\mathcal{B}}}
\def\Ecalv{\bm{\mathcal{E}}}
\def\Ecal{{\mathcal E}}
\def\Bcal{{\mathcal B}}
\def\Fcal{{\mathcal F}}
\def\Acal{{\mathcal A}}
\def\tr{{\rm tr}}

\newcommand{\sgn}{{\mathrm{sgn}}}

\begin{document}
\title{Theory of  diffusive $\varphi_0$ Josephson junctions in the presence of spin-orbit coupling}
\author{F. S. Bergeret}
\affiliation{Centro de F\'{i}sica de Materiales (CFM-MPC), Centro
Mixto CSIC-UPV/EHU and Donostia International Physics Center (DIPC),
Manuel de Lardizabal 4, E-20018 San
Sebasti\'{a}n, Spain}
\author{I. V. Tokatly}
\affiliation{Nano-Bio Spectroscopy group, Departamento de F\'isica de Materiales, Universidad del Pa\'is Vasco, Av. Tolosa 72, E-20018 San Sebasti\'an, Spain}
\affiliation{IKERBASQUE, Basque Foundation for Science, E-48011 Bilbao, Spain}

\begin{abstract}
We present a full microscopic theory based on the SU(2) covariant formulation of the  quasiclassical formalism to describe the Josephson current through an  extended  superconductor-normal metal-superconductor (SNS) diffusive junction with an intrinsic spin-orbit  coupling (SOC)  in the presence of a spin-splitting field ${\bf h}$. We demonstrate that  the ground  state of the junction corresponds to a finite { intrinsic}  phase difference $0<\varphi_0<2\pi$  between the superconductor electrodes provided that both, ${\bf h}$ and the SOC-induced SU(2) Lorentz force are finite.  In the particular case of a Rashba SOC we present analytic and numerical results for  $\varphi_0$ as a function of the strengths of the spin fields, the length of the junction, the temperature and the {properties of} SN interfaces.

\end{abstract}
\maketitle

The dc Josephson effect establishes that the supercurrent flowing between two superconductors connected  by a weak link (normal metal, ferromagnet or semiconductor) is given by
$I_J=I_c\sin \varphi\; $. Here $\varphi$ is the phase difference between the superconducting electrodes and $I_c$ the critical current, {\it i.e.} the maximum supercurrent that can flow through the junction.
The ground state of such a junction corresponds to a zero current state and vanishing phase difference and  for that reason it is denoted as a $0$-junction.
In analogy,  one can define a $\varphi_0$-junction with a more general current phase relation described by 
 \begin{equation}
 \label{JJphi}
I_J=I_c\sin (\varphi+\varphi_0)\;.
\end{equation}
The ground state corresponds to  a finite phase difference $\varphi_0$ across the junction.  
Examples of non-zero junctions are superconductor-ferromagnet-superconductor (SFS) junction of certain thickness with a ground state at $\varphi_0=\pi$, 
predicted {in 1982} \cite{pijunction}  and first detected  in 2001 by Ryazanov {\it et al.}\cite{Ryazanov2001}.  

Besides $0$ and $\pi$ junctions currently  there is no experimental evidence of a $\varphi_0$-junction with $0<\varphi_0<\pi$ for single junctions \cite{note0,Goldobin2011,Sickinger}.
It was, however, theoretically suggested by A. Buzdin \cite{Buzdin2008} that if the weak link is made of 
a  non-centrosymmetric magnetic metal such a  junction is possible. This prediction has been originally formulated 
in terms of  the Ginzburg-Landau (G-L) { theory} 
written in the presence of  a Rashba-like SOC and an exchange field.  $\varphi_0$-junctions have been also analyzed in different types of ballistic junctions with the Rashba SOC \cite{ballistic}.
The universality of this result in the presence of disorder or a generic SOC is however  questionable, 
and no further conclusions can be drawn from previous works.

In this letter we address this question and  present a complete theoretical description of the Josephson effect through a weak link with  arbitrary linear in momentum  SOC and disorder. 
First, by using { the analogy of SOC to a SU(2) gauge field and simple symmetry arguments}
we set the conditions for the  $\varphi_0$-junction to exist in  a ballistic system, and demonstrate the connection between $\varphi_0$-junction behavior  and  the Edelstein effect in superconductors\cite{Edelstein}.   In a second part we use the SU(2) covariant Eilenberger equation\cite{Eilenberger} to show that SOC-induced SU(2) Lorentz force
is responsible for the intrinsic $\varphi_0$  for an arbitrary degree of disorder. Finally we analyze in detail the Josephson current in a diffusive SNS junctions with  Rashba SOC and a spin-splitting field and compute the phase  $\varphi_0$ for a broad range of parameters. 
 
Our starting point is the following Hamiltonian describing a ferromagnetic metal with a linear in momentum SOC
 \begin{equation}
\label{H-linearSO}
H_0 = \frac{1}{2m}(p_j - \hat{\Acal}_j)^2 + \hat{\Acal}_0 +V_{\rm imp}, 
\end{equation}
where  $V_{\rm imp}$ is the random potential of impurities, $\hat{\Acal}_0=\frac{1}{2}\Acal_j^a\sigma^a\equiv h^a\sigma^a$ describes the exchange field of a ferromagnet,  and $\hat{\Acal}_j=\frac{1}{2}\Acal_j^a\sigma^a$ parametrizes a generic SOC ($\sigma^a$ are the Pauli matrices) \cite{note_Rashba}. 
According to  Eq.~(\ref{H-linearSO})  the SOC and the Zeeman coupling \cite{note_zeeman} enter the problem as the space ($\mu=j$) and the time ($\mu=0$) components of an effective background SU(2) { gauge field $\hat{\Acal}_{\mu}$} \cite{Mineev92,Frolich93,Jin2006,Tokatly2008}, { which implies the form invariance of the Hamiltonian under local SU(2) gauge transformations \cite{gauge}.} 

 
The general symmetry origin of $\varphi_0$-junctions can be analyzed at the level of the G-L theory in ballistic structures. It has been recognized in Ref. \cite{Buzdin2008} that the appearance of  $\varphi_0$-junction is ultimately related to the existence of a Lifshitz-type invariant in the free energy  $F_{L}\sim {\bm T} {\bm v}_s\sim T_i\partial_i\varphi$, where ${\bm v}_s$ is the superfluid velocity and $\varphi$ is the phase of the condensate.  Such an invariant requires the existence of a polar vector ${\bm T}$ that is odd under time reversal \cite{toroid}. For a system described by the Hamiltonian (\ref{H-linearSO}) the form of this vector can be uniquely constructed using the SU(2) gauge symmetry arguments. {As the energy must be SU(2) gauge invariant \cite{Tokatly2008},  so has to be the vector ${\bm T}$. Hence,}  the components $\hat{\Acal}_{\mu}$  can enter only via trace of powers of the SU(2) field strength tensor
\begin{equation}
 \label{field-tensor}
 \hat{\Fcal}_{\mu\nu}=\frac{1}{2}\Fcal^a_{\mu\nu}\sigma^a = \partial_{\mu}\hat{\Acal}_{\nu} - \partial_{\nu}\hat{\Acal}_{\mu}
 -i[\hat{\Acal}_{\mu},\hat{\Acal}_{\nu}].
\end{equation}
To the lowest order in SOC the vector { with required properties} is uniquely defined as $T_i\sim\tr(\hat{\Fcal}_{0k}\hat{\Fcal}_{ki})$, thus
\begin{equation}
\label{Lifshitz}
F_L\sim\tr(\hat{\Fcal}_{0k}\hat{\Fcal}_{ki})v_{s,i} = (\Ecalv^a\times\Bcalv^a){\bm v}_s,
\end{equation}
where $\Ecal_k^a={\Fcal}_{0k}^a$ and $\Bcal_i^a=\varepsilon_{ijk}\hat{\Fcal}_{jk}$ are the SU(2) electric and magnetic field vectors, respectively. 
Therefore the $\varphi_0$-junction behavior  requires a configuration of the Zeeman and SO couplings for which a cross product of the effective SU(2) electric and magnetic fields has a component along the Josephson current. In the static case  the SU(2) electric field is given by $\hat{\Fcal}_{0k}=-\tilde{\nabla}_k\hat{\Acal}_0$, where $\tilde{\nabla}_k\cdot=\partial_k\cdot -i [\hat{\Acal}_k,\cdot]$ is the covariant derivative. Therefore the Lifshitz invariant of Eq.~(\ref{Lifshitz}) can be  written as $F_L\sim \tr(\hat{\Acal}_0\tilde{\nabla}_k\hat{\Fcal}_{ki})v_{s,i}$. This representation makes a connection of the $\varphi_0$-junction behavior  to the Edelstein effect in superconductors with SOC \cite{Edelstein}. Indeed,  the relevant contribution to the free energy can be interpreted as a Zeeman coupling of the exchange field to the Edelstein spin density $\delta\hat{S}\sim v_{s,i}\tilde{\nabla}_k\hat{\Fcal}_{ki}$ induced by the supercurrent.   Thus the $\varphi_0$-junction is realized if the Josephson 
current through the link generates (via the SOC)  a spin component parallel to the exchange field. {For the explicit $\Acal_k^a$ of the Rashba SOC \cite{note_Rashba}} we find $F_L\sim\alpha^3({\bm h}\times\hat{\bm z}){\bm v}_s$. Thus, by using only symmetry arguments we recover (up to a numerical factor) the result of lengthy calculations in Ref.~\onlinecite{Edelstein} and show that in general the $\varphi_0$-junction effect is at least cubic in SOC constant.
In the rest of this letter  we  demonstrate that the symmetry-based existence conditions remain
valid even in the diffusive limit although the explicit form of $\varphi_0$ becomes quite different from that predicted by the G-L theory of clean superconductors. 
 
Formally, conditions for the existence of an intrinsic $\varphi_0$ in a  SNS junction can be determined at the level of the Eilenberger  equation  in the ferromagnetic N layer. To obtain this equation we introduce the  gauge covariant quasiclassical Green's functions \cite{Gorini2010,BT2014,Konschelle} and focus on the stationary case, in which the  Eilenberger equation  follows directly from Eq.~(40) of Ref.~\onlinecite{BT2014},
\begin{eqnarray}
&v_Fn_k\tilde\nabla_k \check g+\left[(\omega-i\hat{\Acal}_0)\tau_3,\check g\right]-
\frac{1}{2m}\left\{ \hat{\cal F}_{jk},n_j\frac{\partial}{\partial n_{k}}\check g\right\}=&\nonumber\\
&-\frac{1}{2\tau}\left[\langle\check g\rangle,\check g\right]&\;, 
\label{fullEil}
\end{eqnarray}
where $\check g(\omega,{\bm n},{\bm r})$ is the 4$\times$4 quasiclassical { covariant} Green's function matrix in Nambu-spin space, which depends on the Matsubara frequency $\omega$, the direction ${\bm n}$ of the momentum at the Fermi surface (FS), and the spatial coordinate ${\bm r}$. In Eq.~(\ref{fullEil}) $\tau$ is the momentum relaxation  time due to impurities and $\langle ... \rangle$ stands for the ${\bm n}$-average. The SOC enters this equation via the covariant derivative in the first term, and via the SU(2) magnetic field $\hat{\cal F}_{jk}$ in the last term in the r.h.s. {The former leads to a SOC-induced spin precession, while the latter describes the SU(2) Lorentz force that causes a spin-dependent deflection of trajectories of the FS electrons \cite{Lorentz-force}.} In the normal metals  the SU(2) Lorentz force is the origin of  the spin Hall and the Edelstein effects \cite{Gorini2010,Raimondi2012}.   Below we show  that in superconductors the  SU(2) Lorentz force is also responsible for the intrinsic anomalous phase $\varphi_0$. 

To simplify the further analysis  we assume that either the proximity effect is weak (due to a small barrier transmission between the S electrodes and N), or the temperature $T$ is close to the critical superconducting temperature $T_c$. In this case the anomalous Green's function $\hat{f}$ in the N region is small and the full $\check g$ is approximated as $\check g\approx\left(
\begin{array}{cc}
{\rm sgn}\omega &\hat f\\
-\hat{\bar f}&-{\rm sgn}\omega 
\end{array}
\right)$,
%
where $\hat{\bar f}({\bm n})=\sigma_y\hat{ f}^*(-{\bm n})\sigma_y$ is the time-reversal conjugate anomalous function.
Then Eq.~(\ref{fullEil}) can be linearized:
\begin{eqnarray}
&v_Fn_k\tilde\nabla_k \hat f+\left\{(\omega-i\hat{\Acal}_0),\hat f\right\}-
\frac{1}{2m}\left\{\hat{\cal F}_{jk},n_j\frac{\partial}{\partial n_{k}} \hat f\right\}&\nonumber\\
&=-\frac{{\rm sgn}\omega}{\tau}\left(\hat f-\langle\hat f\rangle\right)&
\label{linEil}
\end{eqnarray}
We first assume that the S/N interfaces have high boundary resistance $R_b$. Thus,  the boundary  conditions (BCs) at the interfaces (located at $x=0,L$) read \cite{BVE2001_Jos}
\begin{equation}
\pm{\rm sgn}\omega\cdot n_x\hat{f}\big\vert_{x=0,L}=\frac{\cal T}{4} f_{BCS} e^{\mp i\frac{\varphi}{2}} \label{bcEil1}
\end{equation}
where $f_{BCS}=\Delta/\sqrt{\omega^2+\Delta^2}$ is the BCS bulk anomalous Green's function in the left and right superconducting leads, $\varphi$ is the given phase difference across the junction, and ${\cal T}$ is the transmission coefficient which for simplicity is assumed to be momentum independent. 
The corresponding  equations for $\hat{\bar f}$ are obtained by applying the operation of time reversal, $\hat{\bar O}({\bm n})=\sigma_y\hat{ O}^*(-{\bm n})\sigma_y$, to  Eqs.~(\ref{linEil})-(\ref{bcEil1}).  
In terms of the anomalous Green's functions the  Josephson current density is given by 
\begin{equation}
j_k=\frac{ie\pi}{2}N_0v_FT\sum_\omega {\rm sgn }\omega\cdot\tr_{\sigma} \langle n_k\hat{\bar f}\; ,
\hat{f}\rangle
\label{currEil}
\end{equation} 
where $N_0$ is the normal density of states at the Fermi level.

A striking signature of the $\varphi_0$-junction is a nonzero anomalous Josephson current $I_J=I_c\sin\varphi_0$ in the absence of the external phase difference, $\varphi=0$. From Eqs.~(\ref{linEil})-(\ref{bcEil1}) and their time-reversal conjugate for $\varphi=0$ one easily finds that if  $\hat{\cal F}_{kj}=0$ then $\hat{\bar f}({\bm n})=\hat{f}(-{\bm n})$ and therefore after averaging over the momentum direction in Eq.~(\ref{currEil}) the Josephson current vanishes. On the other hand, if $\hat{\cal F}_{kj}\ne 0$, but $\hat{\Acal}_0=0$ we find the relation $\hat{\bar f}(\omega)=\hat{f}(-\omega)$ and the current again vanishes upon the frequency summation. Hence a nonzero $\varphi_0$ can appear only due to a simultaneous action of the SOC-induced SU(2) Lorentz force and the exchange field. We now  refine further  this argument for  the diffusive limit, in which the Eilenberger equation reduces to a diffusion-like equation for the anomalous Green's functions, the so called Usadel equation \cite{Usadel}. 

To derive the Usadel equation in the presence of the SU(2) magnetic field we follow the standard route \cite{LO-Book,BT2014}. In the diffusive limit the  relaxation time $\tau$ plays a role of the small parameter. Due to frequent collisions with impurities the Green's function becomes almost isotropic $\hat{f}\approx \hat{f}_0 + n_k\hat{f}_{1,k}$, where $\hat{f}_0 = \langle\hat{f}\rangle$ and the vector $\hat{\bm f}_{1}$ determines the leading anisotropic correction. Inserting this ansatz into Eq.~(\ref{linEil}) and performing expansion in $\tau$ we find
\begin{eqnarray}
 \nonumber
 &\hat{f}_{1,k}=-\sgn\omega\cdot\tau v_F\tilde{\nabla}_k\hat{f}_0 - \frac{\tau^2v_F}{2m}
 \{\hat{\cal F}_{kj},\tilde{\nabla}_j\hat{f}_0\} &\\
& -i\tau^2v_F\{\hat{\cal F}_{0k},\hat{f}_0\}.&\; ,
\label{f1}
\end{eqnarray}
whereas  the isotropic term $\hat{f}_0 = \langle\hat{f}\rangle$  obeys the linearized Usadel equation 
\begin{eqnarray}
&D\tilde\nabla^2 \hat{f}_0 - 2|\omega|\hat{f}_0 +i\sgn\omega
\{\hat{A}_0 + \tau D\tilde{\nabla}_k\hat{\Fcal}_{0k}, \hat{f}_0\}&\nonumber \\
&+ \sgn\omega\cdot\tau D\left\{\frac{1}{2m}\tilde{\nabla}_j\hat{\Fcal}_{jk} 
+ i\hat{\Fcal}_{0k},\tilde{\nabla}_k\hat{f}_0\right\}=0\; .& 
\label{linUsa}
\end{eqnarray}
Here $D$ is the diffusion coefficient. { The Josephson current and the BCs in the diffusive regime are obtained by inserting the form  $\hat{f}\approx \hat{f}_0 + n_k\hat{f}_{1,k}$ into Eqs.~(\ref{currEil}) and (\ref{bcEil1}), respectively, and performing the ${\bm n}$-average.} Notice that  the SU(2) magnetic field enters the Usadel equation of Eq.~(\ref{linUsa}) only in a form of the covariant divergence $\tilde{\nabla}_j\hat{\Fcal}_{jk}$. Importantly, only the term $\sim\tilde{\nabla}_j\hat{\Fcal}_{jk}$ changes sign under time reversal thus making Eq.~(\ref{linUsa}) for $\hat{f}_0$ different from its time-reversal counterpart for $\hat{\bar f}_0$.  If SOC is such that $\tilde{\nabla}_j\hat{\Fcal}_{jk}=0$ then for $\varphi=0$ we have $\hat{\bar f}_0=\hat{f}_0$ which implies vanishing anomalous Josephson current and $\varphi_0=0$ or, possibly,  $\varphi_0=\pi$. A closer inspection of Eq.~(\ref{linUsa}) shows that we can get a nontrivial $\varphi_0$ only if $T_k=\tr(\hat{\Acal}_0\tilde{\nabla}_j\hat{\Fcal}_{jk})$ has a nonzero component in the direction of spatial inhomogeneity (the direction of the current). This fully agrees with the analysis of the Lifshitz invariant in the free energy. It is worth noting that the quantity $\tilde{\nabla}_j\hat{\Fcal}_{jk}$ 
determines the equilibrium spin current $J_k^a$ in the normal state \cite{Tokatly2008}.  Hence the condition for the existence of $\varphi_0$-junction can be restated as $h^aJ_k^a\ne 0$.

Equation (\ref{linUsa}) describes the condensate in a diffusive normal region with arbitrary (possible inhomogeneous) SOC and  spin-splitting field. Here,  as a specific example  we consider a SNS lateral  junction with  a Rashba SOC corresponding to ${\cal A}_x^y=\alpha$ and ${\cal A}_y^x=-\alpha$.  Since for this configuration of SO fields the interesting effects are proportional only to the component of $\hat{\Acal}_0$ perpendicular to the current direction $x$, we assume that $\hat{\Acal}_0=h\sigma_y$. Thus, the general solution of Eq.~(\ref{linUsa}) { has the form $\hat{f}_0=f_s-i{\rm sgn}\omega f_t \sigma^y$,}
where $f_s$ and $f_t$ are the singlet and triplet components of the condensate \cite{BVE_rmp}. The general Usadel equation simplifies as follows
 \begin{eqnarray}
\partial_{x}^2 f_s-\kappa_\omega^2f_s+\kappa_h^2 f_t-2i\kappa_\alpha\partial_x f_t&=&0\label{usadel}\\
\partial_{x}^2 f_t-(\kappa_\omega^2+\alpha^2)f_t-\kappa_h^2 f_s+2i\kappa_\alpha\partial_x f_s&=&0\label{usadelb}\; ,
\end{eqnarray}  
where  $\kappa_\omega^2=2|\omega|/D$, $\kappa_h^2=2h/D$, $\kappa_\alpha=\tau\alpha^3/4m$, and we have neglected terms of the order $(\alpha l)^2$.  
{ The term $\alpha^2$ in Eq.~(\ref{usadelb})  comes from the covariant Laplacian $\nabla^2 \hat{f}_0$ in Eq.~(\ref{linUsa}) and describes the Dyakonov-Perel (DP) relaxation of the triplet component \cite{BT2014}.}  The effect of the SU(2) Lorentz force is encoded in the parameter $\kappa_\alpha$.  Notice that a structurally similar equations have been considered in Ref.\cite{Malshukov} in the context of the spin-hall effect.

Equations (\ref{usadel}), (\ref{usadelb}) have to be complemented with BCs at the interfaces with the superconducting leads at $x=0,L$. 
We analyze  two type of interfaces: (i)   those with a finite resistance $R_b$ described by the generalized Kupriyanov-Lukichev BCs \cite{KL} which follow from Eq.~(\ref{bcEil1}) 
\begin{equation}
[\partial_x f_s-i\kappa_\alpha f_t]_{0,L}=\pm\gamma f_{BCS}e^{\mp i\varphi/2}, \quad
\partial_x f_t\vert_{0,L}=0
\label{BCt}
\end{equation}
and (ii) those fully transparent for which {the Green's functions are continuous at the interfaces, that is 
$\hat{f}_0\vert_{x=0,L} = f_{BCS}e^{\mp i\phi/2}$.}
Eqs.~(\ref{usadel})-(\ref{usadelb}) together with the BCs determines completely the anomalous Green function in the N region in the presence of SOC. In terms of $f_s$ and $f_t$ the Josephson current is given by 
\begin{equation}
\label{current}
j=\frac{\pi\sigma_N}{e}T\sum_\omega{\rm Im }\left\{ f_s^*\left(\partial_xf_s-2i\kappa_\alpha f_t\right)-f_t^*\partial_xf_t\right\}\; ,
\end{equation} 
where $\sigma_N=2e^2N_0D$ is the conductivity in the normal state.

\begin{figure}[t]
\includegraphics[width=\columnwidth]{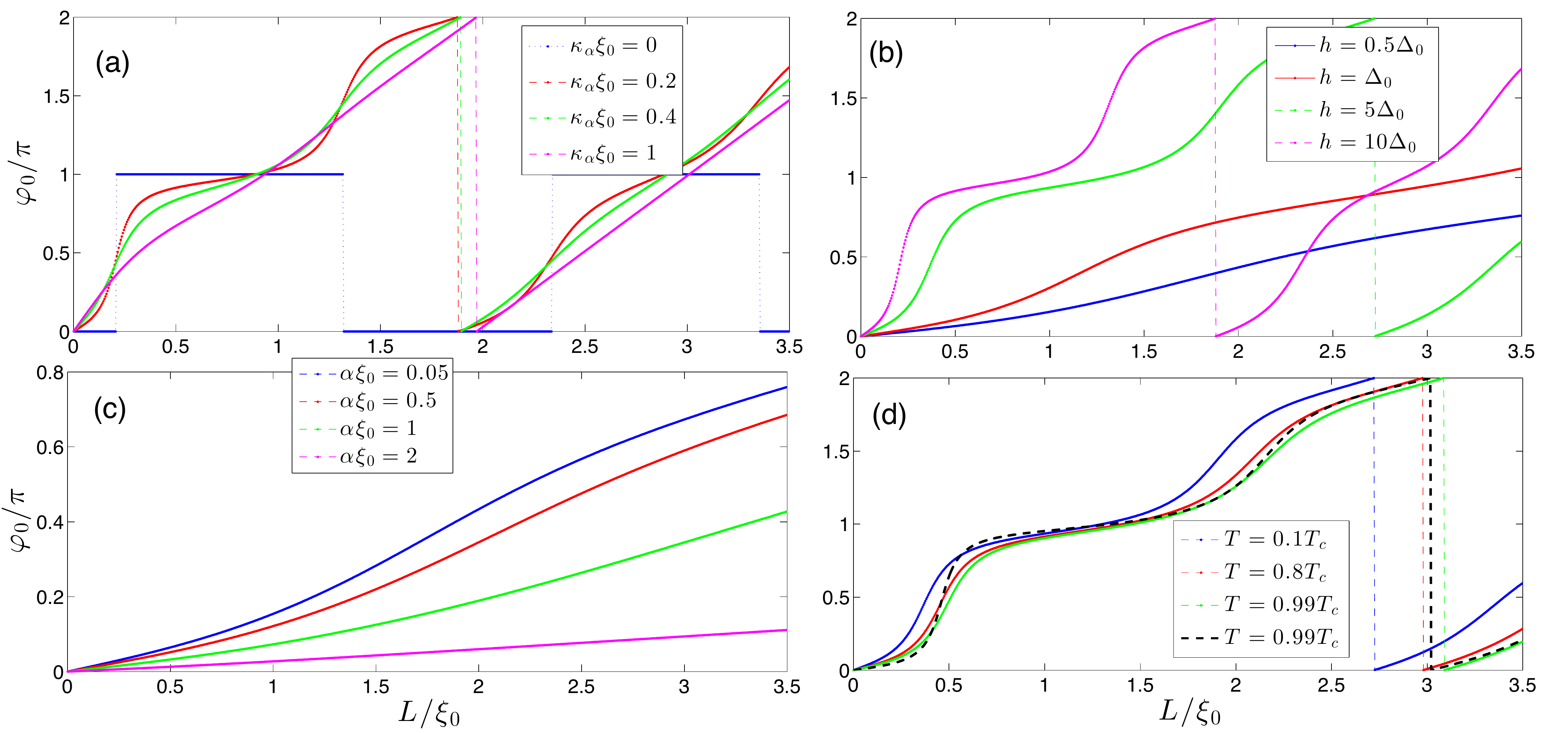}
\caption{\label{fig2}  The dependence of $\varphi_0$ on the junction length $L$ for
(a) different values of  $\kappa_\alpha$, $h=10\Delta_0$, $\alpha \xi_0=0.05$;  (b) different values of $h$, $\kappa_\alpha\xi_0=0.2$, $T=0.1T_c$ and $\alpha \xi_0=0.05$; (c) different values of $\alpha$, $h=0.5\Delta_0$, $\kappa_\alpha\xi_0=0.2$ and $T=0.1T_c$; (d) different values of $T$, $h=5\Delta_0$, $\kappa_\alpha\xi_0=0.2$ and $\alpha \xi_0=0.05$.  The dashed black line in  panel (d) is the high temperature approach obtained by taking only the first term of the sums in Eq. (\ref{ana1}) }
\end{figure}
In principle the boundary problem defined by  Eqs. (\ref{usadel})-(\ref{usadelb})  and the corresponding BCs is linear and can be solved analytically.  
One can demonstrate by using  Eq. (\ref{current}) that the current is given by by Eq. (\ref{JJphi}).
Here we present compact expressions for $\varphi_0$ that  are obtained in certain limiting cases.  
{ For a weak SOC we  disregard the DP term $\alpha^2$, and assume that $\kappa_\alpha\ll|\kappa|\equiv |\sqrt{i\kappa_h^2+\kappa_\omega^2}|$. Furthermore, assuming that the temperature is large enough  ($T\gg\Delta(T)$) we keep in Eq. (\ref{current})  only the contribution of the lowest Matsubara frequency  ($\omega=\pi T$). Then, using BCs of Eq.~(\ref{BCt}), which corresponds to  SN interface  with a finite resistance $R_b$, one obtains}
\begin{equation}
\varphi_0^{R_b}\approx\arctan\left\{\tanh(\kappa_\alpha L)\frac{{\rm Im}\left[\kappa\sinh(\kappa L)\right]}{{\rm Re}\left[\kappa \sinh(\kappa L)\right]}\right\}\label{ana1}
\end{equation}
{ In the case of fully transparent interfaces (zero SN resistance)  the anomalous phase takes the form }
\begin{equation}
\varphi_0^0\approx\arctan\left\{\tanh(\kappa_\alpha L) \frac{{\rm Im}\left[{\kappa^*}\sinh(\kappa L)\right]}{{\rm Re}\left[{\kappa^*}\sinh({\kappa}L)\right]}\right\}\label{ana2}
\end{equation}
{To the lowest order in $h$ Eqs.~(\ref{ana1}),(\ref{ana2}) simplify as follows}
\begin{equation}
\varphi_0^{R_b,0}\approx \frac{\kappa_h^2}{2\kappa_T^2}\tanh(\kappa_\alpha L)\left[\frac{\kappa_T L}{\tanh(\kappa_T L)}\pm 1\right]\; ,
\end{equation}
where $\kappa_T=\sqrt{2\pi T/D}$. From this equation one can see that in the long junction limit ($\kappa_T L\gg 1$) $\varphi_0$ is the same in both cases.
In the opposite limit ($\kappa_T L\ll1$)  the result depends strongly on the type of interface: While for a finite $R_b$ $\phi_0^{R_b}\approx(\kappa_h^2/\kappa_T^2)\kappa_\alpha L$, for the fully transparent interface $\varphi_0^0\approx\kappa_h^2\kappa_\alpha L^3/6$.  Thus in short junctions the effect is much weaker for transparent interfaces.

Now  we turn to the full numerical solution of the boundary problem  and determine the phase $\varphi_0$.  Fig. \ref{fig2} shows the length dependence of $\varphi_0$ for junctions with finite $R_b$.   Panel (a) clearly shows that in the absence of SOC ($\kappa_\alpha=0$, blue horizontal lines), $\varphi_0$ can only equals to $0$ and $\pi$, as it is well known from the theory of SFS junctions\cite{pijunction}.  However, for finite values of $\kappa_\alpha$,  $\varphi_0$  can range between $0$ and $2\pi$. 
Panel (b)  demonstrates that the range of possible values  of  $\varphi_0(L)$ increases with increasing the value of $h$.
From panels (a) and (b) we confirm that both $h$ and $\kappa_\alpha$, have to be finite in order to get a $\varphi_0$-junction.
The effect of the DP relaxation term ($\alpha$)  on $\varphi_0$ is analyzed in Fig. \ref{fig2}c.   It is known that large enough extrinsic SOC suppresses the triplet correlations \cite{BVE_rmp}. This explains the suppression of $\varphi_0$ towards $0$ by increasing $\alpha$ (see Fig. \ref{fig2}c).  Finally, panel (d) in Fig. \ref{fig2} shows  a weak variation of the $\varphi_0(L)$ by varying the temperature.

\begin{figure}[t]
\includegraphics[width=\columnwidth]{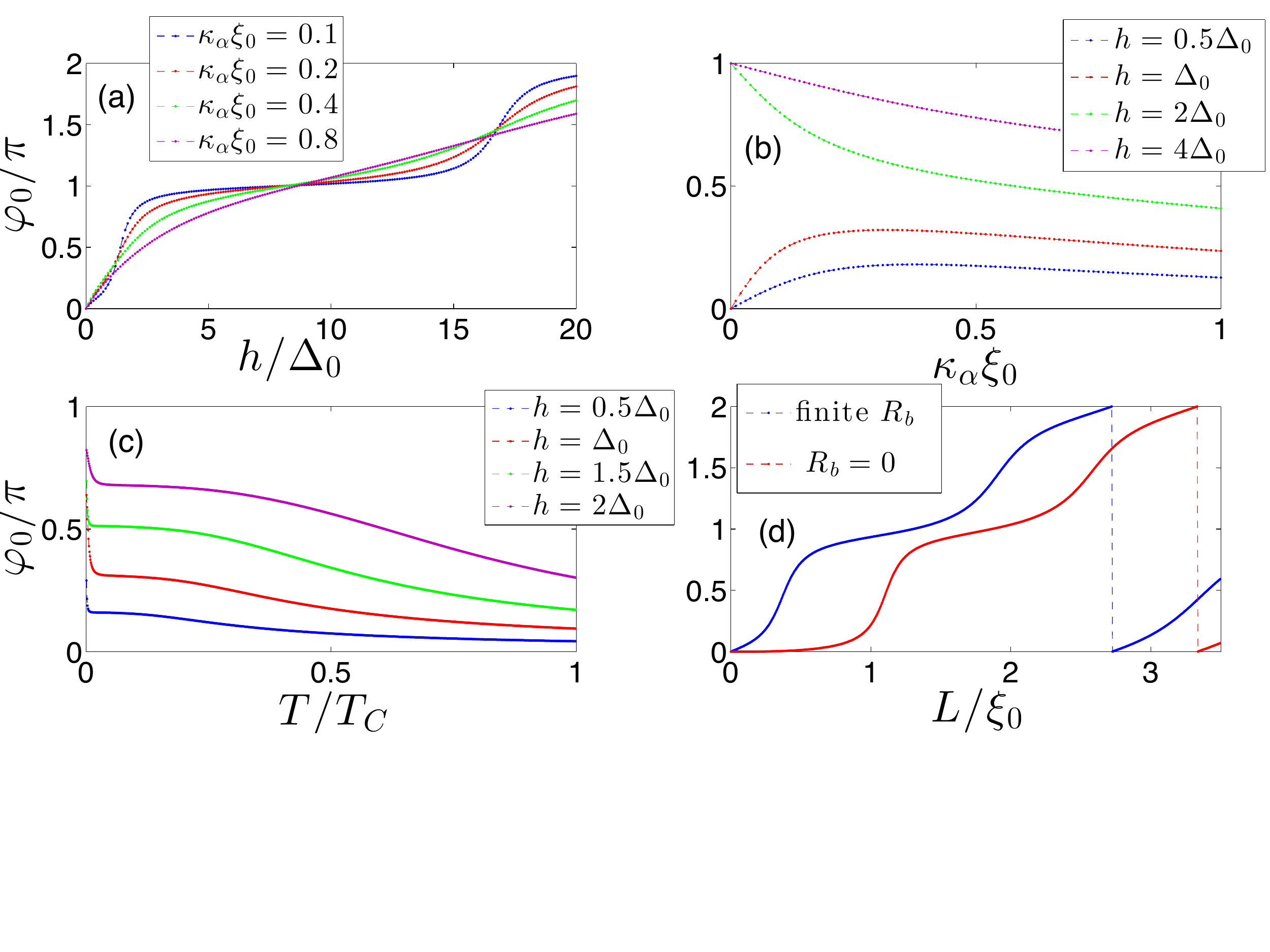}
\vspace{-20mm}
\caption{\label{fig3} 
The dependence of $\varphi_0$ on: 
(a) $h$ for different $\kappa_\alpha$, $L=\xi_0$ and  $T=0.1T_c$ ;
(b) $\kappa_\alpha$ for different $h$, $L=\xi_0$ and  $T=0.1T_c$; (c) $T$ for different $h$, $L=\xi_0$ and  $\kappa_\alpha\xi_0=0.2$ . 
Panel (d) shows the $\varphi_0$ dependence on $L$ for a finite barrier resistance between the S electrodes and the N bridge and for a fully transparent contact ($R_b=0$).  In all panels $\alpha \xi_0=0.05$.}
\end{figure}

The spin-splitting field can be either 
the intrinsic exchange field of a ferromagnet or can be induced by applying an external magnetic field. In the latter case one could tune the value of $\varphi_0$ by measn of an external magnetic field, as shown in Fig.\ref{fig3}a. It is interesting to note that in the presence of a finite $\kappa_\alpha$ the switch on of a magnetic field leads to an enhancement of the current through the junction. If $\kappa_\alpha=0$  only the  $\varphi_0=0,\pi$ are possible. The full dependence of $\varphi_0$ on $\kappa_\alpha$ is shown in Fig. \ref{fig3}c for different values of the spin-splitting field $h$.  The value of $\varphi_0$ can also be tuned by changing the temperature as shown in Fig.\ref{fig3}. 
Finally, the dependence  $\varphi_0(L)$ in the case of a finite and vanishing interface resistance are shown in Fig.\ref{fig3}d. 
Although the curves are qualitatively very similar there is a clear quantitative difference, specially in short junctions in accordance with the analytic expressions obtained above. 

In conclusion, {we presented a SU(2) covariant theory, which} 
allows for the full description of the $\varphi_0$-junction  behavior  in SNS structures with linear in momentum SOC and a spin-splitting field. Simple symmetry arguments  at the level of the covariant Eilenberger equation, 
show that a finite phase $\varphi_0$  [see Eq. (\ref{JJphi})], with $0<\varphi_0<\pi$, can appear due to a simultaneous action of the SOC-induced SU(2) Lorentz force and the spin-splitting field, independently of the degree of disorder. In particular, we have computed the Josephson current in a diffusive SNS structure,  and set the conditions for the $\varphi_0$-junction behavior as a function of different parameters. We finally demonstrated that for short junctions a finite resistance between the S and the N leads to a larger value of $\varphi_0$.

{\it Acknowledgments} We thank F. Konschelle for useful discussions. This  work  was supported by the Spanish Ministry of Economy
and Competitiveness under Projects No. FIS2011-28851-C02- 02, and FIS2013-46159-C3-1-P,  the Basque Government under UPV/EHU Project No. IT-756-13.
 I.V.T. acknowledges funding by the Grupos Consolidados UPV/EHU del Gobierno Vasco (Gant No. IT578-13). 
F.S.B thanks Martin Holthaus and his group for their kind hospitality at the Physics Institute of the Oldenburg University.


\begin{thebibliography}{30}

\bibitem{pijunction}  A. I. Buzdin, L. N. Bulaevskii, and S.V. Panjukov, JETP
Lett. 35, 178 (1982).

\bibitem{Ryazanov2001} V. V. Ryazanov, V. A. Oboznov, A. Yu. Rusanov, A. V. Veretennikov, A. A. Golubov, and J. Aarts,  Phys. Rev. Lett. {\bf 86}, 2427 (2001).

\bibitem{note0} The $\varphi_0$ behavior can be  realized in a hybrid configuration of $0$- and $\pi$-junctions in parallel  as shown in Refs.\cite{Goldobin2011, Sickinger}.

\bibitem{Goldobin2011} E. Goldobin, D. Koelle, R. Kleiner, and R. G. Mints, Phys. Rev. Lett. {\bf 107}, 227001 (2011).

\bibitem{Sickinger} Sickinger, A. Lipman, M. Weides, R. G. Mints,
H. Kohlstedt, D. Koelle, R. Kleiner, and E. Goldobin, Phys.  Rev. Lett. {\bf 109}, 107002 (2012).

\bibitem{Buzdin2008} A. I. Buzdin, Physical Review Letters 101, 107005 (2008),

\bibitem{ballistic} A. A. Reynoso, G. Usaj, C. Balseiro, D. Feinberg, and M. Avignon, Phys.  Rev.  Lett. {\bf 101}, 107001 (2008);  
A. Zazunov, R. Egger, T. Martin, and T. Jonckheere, Phys. Rev. Lett. {\bf 103}, 147004 (2009);  J.-F. Liu and K. Chan, Phys.  Rev. B {\bf 82}, 125305 (2010); 
T. Yokoyama, M. Eto, and Y. V. Nazarov, Jour. of the Phys. Soc.  of Japan{\bf  82}, 054703 (2013).

\bibitem{Edelstein} V. M. Edelstein, Phys. Rev. Lett. {\bf 75}, 2004 (1995).

\bibitem{Eilenberger} G. Eilenberger, Z. Phys. 214, 195 (1968).



\bibitem{note_Rashba} { A particular case of the Rashba  SOC\cite{Rashba}, $H_R=\frac{\alpha}{m}({\bm p}\times\hat{\bm z}){\bm\sigma}$, corresponds to ${\cal A}_x^y=\alpha$ and ${\cal A}_y^x=-\alpha$ and all other components are zero.}

\bibitem{note_zeeman} The  origin of the Zeeman spin-splitting field in the Hamiltonian Eq.~(\ref{H-linearSO}) can be either an external magnetic field applied in the plane of the N bridge, or the  intrinsic exchange field of a ferromagnet in a SFS junction.

\bibitem{Mineev92}  V. P. Mineev and G. E. Volovik, Journal of Low
Temperature Physics \textbf{89}, 823 (1992).

\bibitem{Frolich93}  J. Fr\"{o}hlich and U. M. Studer, Rev. Mod Phys. 
\textbf{65}, 733 (1993).

\bibitem{Tokatly2008} I. V. Tokatly, Phys. Rev. Lett. {\bf 101}, 106601 (2008).

\bibitem{Jin2006}  P.-Q. Jin, Y.-Q. Li, and F.-C. Zhang, J. Phys. A: Math. Gen. {\bf 39}, 7115 (2006).

\bibitem{gauge} { The SU(2) gauge transformation corresponds to a local rotation of spinor wave functions $\psi\mapsto \hat{U}\psi$ with a SU(2) matrix $\hat{U}$, supplemented with the transformation of the potential $\hat{\Acal}_\mu\mapsto\hat{U}\hat{\Acal}_\mu\hat{U}^{-1} -i(\partial_\mu\hat{U})\hat{U}^{-1}$.}



\bibitem{toroid} It is worth noting that a $t$-odd polar vector has a symmetry of a toroid moment \cite{Dubovik}. The ``phase generation'' effect in a hypothetical  superconductor with a toroidal ordering was discussed long ago in ref. \cite{Gorbatsevich}


\bibitem{BT2014} F. S. Bergeret, and I. V. Tokatly, Phys. Rev. B {89},134517 (2014).

\bibitem{Konschelle} Fran\c{c}ois Konschelle,  Eur. Phys. Jour. B {\bf 87}, 119  (2014); {\it ibid} arXiv:1403.1797 (unpublished).

\bibitem{Gorini2010}  C. Gorini, P. Schwab, R. Raimondi, and A. L. Shelankov, Phys.
Rev. B {\bf 82}, 195316 (2010).

\bibitem{Lorentz-force} Since $\hat{\cal F}_{jk}=-\hat{\cal F}_{kj}$, the Lorentz force operator $\hat{\cal F}_{jk}n_j\frac{\partial}{\partial n_{k}}$ acts only on the direction ${\bm n}$ of momentum. Alternatively it can be written as $\hat{\cal F}_{jk}n_j\frac{\partial}{\partial n_{k}}=\hat{\Bcalv}\frac{\partial}{\partial{\bm\phi}}$, where ${\bm\phi}$ is the polar angle about the direction of the magnetic field.


\bibitem{Raimondi2012} R. Raimondi, P. Schwab, C. Gorini,
and G. Vignale, Ann. Phys. (Berlin) {\bf 524}, 153 (2012).


\bibitem{Usadel} K. L. Usadel, Phys. Rev. Lett. 25, 507 (1970).

\bibitem{LO-Book} A. I. Larkin and
Superconductivity (Elsevier, Amsterdam, 1984).

\bibitem{BVE_rmp} F. S. Bergeret, A. F. Volkov, and K. B. Efetov, Rev. Mod. Phys. 77, 1321 (2005).

\bibitem{Malshukov} A. G. Mal'shukov and C. S.  Chu, Phys. Rev. B {\bf 78}, 104503 (2008).

\bibitem{KL} M. Yu. Kuprianov and V. F. Lukichev, Sov. Phys. JETP {\bf 67}, 1163 (1986).


\bibitem{Rashba} Y. A. Bychkov and E. I. Rashba, PisÕma Zh. Eksp. Teor. Fiz. {\bf 39}, 66 (1984) [JETP Lett. {\bf 39}, 78 (1984)].


\bibitem{Dubovik} V. M. Dubovik, and V. V. Tugushev,  Phys. Rep. {\bf 187} , 145 (1990).

\bibitem{Gorbatsevich} A. A. Gorbatsevich, Sov. Phys. JETP {\bf 68}, 847 (1989).

\end{thebibliography}
\end{document}